\documentclass[a4paper,11pt]{article}

\RequirePackage{amsmath,amssymb,amsfonts,amsthm}
\RequirePackage[numbers]{natbib}

\theoremstyle{plain}
\newtheorem{theorem}{Theorem}[section]

\newtheorem{lemma}[theorem]{Lemma}

\theoremstyle{definition}

\newtheorem{example}[theorem]{Example}

\begin{document}

\begin{center}

  \Large
  {\bf  Scaling Limits of Time-averaged Distributions
        for Symmetric Quantum Walks on Hypercubes}
  \normalsize

  \bigskip

  By

  \bigskip

  \textsc{Yusuke Ide, Norio Konno, Shuhei Mano}


   

\end{center}

\small

{\bf Abstract.}
We study continuous-time and discrete-time quantum walks on
the hypercube $\{0,1\}^d$ associated with symmetric random
walks, that is, random walks whose transition probabilities
depend only on the Hamming distance, using the Hamming scheme
and Krawtchouk polynomials. For continuous-time walks, we
show that the time-averaged distribution of the Hamming
distance is a discrete arcsine law whenever the eigenvalues
are distinct. For discrete-time walks, exact formulas for
the time-averaged distribution are known as combinatorial
sums or as mixtures involving Krawtchouk polynomials, but
their scaling limits as $d\to\infty$ (time average first)
have not been specified. We obtain them explicitly for
three random walks: the simple random walk (an arcsine law),
the random walk with binomial weights (an equal mixture of
arcsine laws with different supports), and the independent
random walk (atoms of mass $1/2$ at $0$ and at $1/2$).
Comparing with the continuous-time limits, the two models
agree for the simple random walk but differ for the other
two, so the correspondence between them is not universal.

\smallskip



\normalsize

\section{Introduction}

Quantum walks are motivated by search algorithms in quantum
computing. Although they are not Markov chains, they
incorporate randomness in a quantum mechanical sense. There
are two distinct classes of such models, discrete-time
(coined) quantum walks and continuous-time quantum walks.
Both evolve by unitary operators, yet they are quite
different models. The discrete-time model requires,
in addition to the position space familiar from Markov chains,
an internal coin space. See, for example, \cite{Kem03,Kon08,Ven12,Por13}.

On $\mathbb{Z}$, the two models already show different limit
behavior. For the Hadamard walk, $X^{(D)}_t/t$ converges
weakly to a density, so-called Konno's distribution
\cite{Kon02} as $t\to\infty$, whereas for
the continuous-time walk, $X^{(C)}_t/t$ converges to
an arcsine law \cite{Kon05}. Both limit densities
have two maxima at the endpoints of their supports, yet
differ, including the supports. These are limits
as time grows on an infinite graph. On finite graphs, a natural
alternative is to compare the time-averaged distributions as
the size of the graph grows. Ide \cite{Ide23} showed that, for
path graphs with a spectral gap, the scaling limits of
the time-averaged distributions of the continuous-time walk and
of the discrete-time walk given by Szegedy’s walk coincide with
the number of vertices grows.
 
This raises the question of when the two models have the same
scaling limit. We address it for walks on the hypercube
$\{0,1\}^d$ associated with symmetric random walks, that is,
random walks whose transition probabilities depend only on
the Hamming distance between vertices. Note that the term
``symmetric'' is used here in a broader sense than is typical
for Markov chains; specifically, transitions are not restricted
to adjacent vertices. For such a random walk,
we consider the continuous-time quantum walk and
the discrete-time quantum walk given by Szegedy’s walk.
Throughout, we first take the time average ($T\to\infty$) for
fixed $d$, and then take the scaling limit $d\to\infty$ of
the distribution of the Hamming distance from the origin
normalised by $d$.

For the continuous-time model associated with the simple random
walk, the time-averaged distribution is known to be a discrete
arcsine law on the Hamming distances \cite{Kon08}.
For the discrete-time model, the time-averaged distribution has
been written as a combinatorial sum \cite{MPAD08}, and as
an equal mixture of a discrete arcsine law and a second distribution
expressed through Krawtchouk polynomials \cite{HIKST18, GM26}.
These representations are exact,
however, to our knowledge no scaling limit has been obtained since
the formula of \cite{MPAD08}. In this paper we determine
the scaling limits in closed form for three symmetric quantum walks;
they turn out to be explicit mixtures of arcsine laws and atoms.

Our contributions are as follows. We prove that, for
the continuous-time quantum walk associated with
any symmetric random walk on the hypercube, the time-averaged
distribution of the Hamming distance is a discrete arcsine law,
provided that the eigenvalues are distinct. We determine explicitly
the scaling limits of the time-averaged distributions of
the discrete-time quantum walks associated with three random walks:
the simple random walk, the random walk with binomial weights of
\cite{GM26}, and the independent random walk (the random walk on
the complete graph). For the first, the limit is the arcsine law
on $(0,1)$. For the second, it is the equal mixture of the arcsine
law on $(0,1)$ and the law of $r+(1-2r)\eta$, where $\eta$
independently follows the arcsine law and $r\in(0,1)$
is the parameter of the binomial
weights. For the third, it is the distribution with mass $1/2$ at
$0$ and mass $1/2$ at $1/2$. Comparing the two models, we find that
the correspondence between them holds in one case and fails in
the other two. For the simple random walk, both models have
the arcsine law on $(0,1)$ as their scaling limit. For
the binomially weighted walk, the continuous-time limit is again
the arcsine law, whereas the discrete-time limit is
the mixture above, so the two limits differ. The independent random
walk has repeated eigenvalues and lies outside the scope of
the above result; its continuous-time limit is the point mass at
the origin \cite{Kon08}, which differs from the discrete-time limit.

Since the continuous-time limit is always the arcsine law on $(0,1)$
when the eigenvalues are distinct, the variety of limits comes entirely
from the discrete-time side. In contrast to the path-graph result
of \cite{Ide23}, where the two models share the same weak limit,
these examples show that agreement between discrete-time and
continuous-time quantum walks is not universal.

The paper is organised as follows. Section~\ref{sec:prelim}
recalls the Hamming scheme and the Krawtchouk polynomials and
defines the class of distance-invariant random walks, together
with the three examples above. Section~\ref{sec:CTQW} treats
continuous-time quantum walks and proves a discrete arcsine
law under the distinct-eigenvalue assumption. Section~\ref{sec:limits}
treats discrete-time quantum walks started at the origin
and derives the scaling limits for each of the three examples.
Section~\ref{sec:disc} discusses the correspondence between
the two models and open problems, such as characterising when
the scaling limits coincide.

\section{Preliminary}\label{sec:prelim}

\subsection{Hamming scheme and Krawtchouk polynomials}
  
Let a state space $X=\{0,1\}^d$ where $d\ge 3$.
Set $\partial(x,y)=|\{i:x_i\neq y_i\}|$ for
$x=(x_i)$, $y=(y_i)$ $\in X$. The distance $\partial$
is called the {\it Hamming distance} and induces a relation
$X\times X$ by $(x,y)\in R_i$ $\Leftrightarrow$ $\partial(x,y)=i$,
$i\in\{0,1,\ldots,d\}$. We denote $|x|\equiv\partial(0,x)$.
The undirected graph $(X,R_1)$ with vertices $X$ and edges $R_1$
is called the {\it hypercube}. A pair of a finite set and
a relation satisfying certain conditions is called
an association scheme, and $(X,\{R_i\}_{i\in\{0,1,\ldots,d\}})$
introduced here is specifically called a Hamming scheme.
The advantage of formulating a problem using an association
scheme is that it can reduce the need to perform direct matrix
calculations. See \cite[Chapter~2]{BBIT21} for details.
The adjacency matrix $A_i$ is defined by
\[
  (A_i)_{x,y}=\left\{
    \begin{array}{ccc}
      1 & \text{if} & (x,y)\in R_i,\\ 
      0 & \text{if} & (x,y)\notin R_i.
    \end{array}\right.
\]
Let $\mathcal{A}$ be the vector subspace of the matrices
$M_X(\mathbb{C})$ spanned by $A_0$, $A_1,\ldots,A_d$.
Here, $\mathcal{A}$ is commutative with the matrix product,
and called a Bose--Mesner algebra. Set
\[
  \kappa_i=|\{y\in X:\partial(x,y)=i\}|=\begin{pmatrix}d\\i\end{pmatrix},
\]
where $\kappa_1$ is the degree of each vertex. Commutative
$A_0$, $A_1,\ldots,A_d$ are simultaneously diagonalised by
primitive idempotents $E_0$, $E_1,\ldots,E_d$ in $\mathcal{A}$
satisfying $E_iE_j=E_i\delta_{i,j}$ and
$E_0+E_1+\cdots+E_d=I$, where $A_0$ is the identity matrix
$I$ and $E_0$ denotes the matrix whose entries are all $1/2^d$.
The base change determines the coefficients
\begin{equation}\label{eigenmat}
  A_i={\sum_{j=0}^d} K_i(j)E_j, \quad
  2^dE_i=\sum_{j=0}^d K_i(j)A_j,
\end{equation}
where $K_i(j)$ are the Krawtchouk polynomials
\begin{equation*}\label{Kpol}
  K_i(j;n,d)=\sum_{l=\max\{0,i+j-d\}}^{\min\{i,j\}}(-1)^l
  (n-1)^{i-l}\begin{pmatrix}j\\l\end{pmatrix}
  \begin{pmatrix}d-j\\i-l\end{pmatrix}
\end{equation*}
with $n=2$.

The generating function for the Krawtchouk polynomials is
\begin{equation}\label{gen}
  \sum_{l=0}^dK_l(j;n,d)s^l=\{1+(n-1)s\}^{d-j}(1-s)^j.
\end{equation}

The orthogonality is with respect to the Binomial distribution:
\begin{equation}\label{orth}
  \sum_{l=0}^dK_i(l;n,d)K_j(l;n,d)
  \left(\begin{array}{c}d\\l\end{array}\right)
    \left(1-\frac{1}{n}\right)^l\left(\frac{1}{n}\right)^{d-l}
    =\delta_{i,j}(n-1)^i
    \begin{pmatrix}d\\i\end{pmatrix}.
\end{equation}



We have a sum formula \cite[Theorem~3.1.3]{FF96}
\begin{equation}\label{sum}
\sum_{i=0}^d\{K_i(j)\}^2=
\frac{\begin{pmatrix}2(d-j)\\d-j\end{pmatrix}
        \begin{pmatrix}2j\\j\end{pmatrix}}
         {\begin{pmatrix}d\\j\end{pmatrix}}.
\end{equation}

\subsection{Symmetric random walks on hypercubes}

Suppose the transition probability matrix of a
discrete-time random walk on the state space $X$
satisfies a symmetry, that is, $P$ is constant on
each orbit $\{R_i\}_{i\in\{0,1,\ldots,d\}}$:
\begin{equation}\label{assump}
  \partial(x,y)=\partial(x',y') \quad \Rightarrow
  \quad (P)_{x,y}=(P)_{x',y'},
\end{equation}
or equivalently that $P$ belongs to Bose--Mesner algebra
$\mathcal{A}$. Then, the transition probability takes
the form of
\begin{equation}\label{trans}
  P=\sum_{i=0}^d\frac{w_i}{\kappa_i}A_i, \quad 
  \text{where} \quad w_i\ge 0,~\sum_{i=0}^d w_i=1.
\end{equation}
The $t$-step transition probability $(P^t)_{x,y}$ for
$(x,y)\in R_h$ becomes \cite{Hor97}
\begin{equation}\label{spec}
  P_t(h)=\frac{1}{2^d}\sum_{i=0}^d\rho_i^t K_i(h),
  \quad h\in\{0,1,\ldots,d\}
\end{equation}
with eigenvalues
\begin{equation}\label{eigen} 
  \rho_i=\sum_{j=0}^d \frac{w_j}{\kappa_j} K_j(i),
  \quad i\in\{0,1,\ldots,d\},
\end{equation}
where $\rho_0=1$ and $|\rho_i|\le 1$, $i\in\{1,\ldots,d\}$.
If a random walk is irreducible and aperiodic, $-1$ is not
an eigenvalue. Since the transition probability \eqref{trans}
is symmetric, the stationary distribution is uniform.
We consider a random walk starting from a state of $X$.
Without loss of generality, let the state be $0$.
In the standard basis of the vector space $\mathbb{C}^X$,
the element corresponding to $0$ is denoted by $e_0$.
Then, at $t=1$,
\begin{equation}\label{trans_1}
  P_{0,x}=e_0P=\sum_{i=0}^d \frac{w_i}{\kappa_i}(e_0A_i)_x
  =\sum_{i=0}^d \frac{w_i}{\kappa_i}\delta_{|x|,i}
  =\frac{w_{|x|}}{\kappa_{|x|}}.
\end{equation}

Some examples of random walks on the hypercube satisfying
\eqref{assump} are:

\begin{example}[The simple random walk]\label{exa:simple}
  A walker at $x$ moves to a neighbour $y\in X$ satisfying
  $\partial(x,y)=1$ with equal probability.
  \[
    w_i=\delta_{i,1}, \quad \rho_i=1-\frac{2i}{d},
    \quad i\in\{0,1,\ldots,d\}.
  \]
  This random walk is irreducible and periodic.
  The spectral gap, $2/d$, vanishes as $d\to\infty$.
\end{example}  

\begin{example}[The independent random walk]\label{exa:indep}
  A walker at $x\in X$ moves to any $y\in X$ with equal
  probability. 
  \[
    w_i=
    \begin{pmatrix}d\\i\end{pmatrix}\frac{1}{2^d},
    \quad \rho_i=\delta_{i,0}, \quad i\in\{0,1,\ldots,d\}.
  \]
  This random walk is irreducible and aperiodic. This random
  walk mixes in exactly one step.
  Note that this random walk coincides with the simple random
  walk on the complete graph with $2^d$ vertices.
\end{example}  

\begin{example}[The random walk with binomially weighted transitions]
  \label{exa:bin}
  A walker at $x\in X$ moves to $y\in X$ satisfying
  $\partial(x,y)=i$ with equal probability.
  For $r\in(0,1)\setminus\{1/2\}$,
  \[
    w_i=\begin{pmatrix}d\\i\end{pmatrix}r^i(1-r)^{d-i},
    \quad \rho_i=\left(1-2r\right)^i, \quad
    i\in\{0,1,\ldots,d\}.
  \]
  This random walk is irreducible and aperiodic. This random walk
  reduces to the independent random walk if $r=1/2$.
  The spectral gap, $2r$, does not vanish as $d\to\infty$.
  The is a version of a random walk considered in
  \cite[Example~3]{CG21}. 
\end{example}

\section{Symmetric continuous-time quantum walks}\label{sec:CTQW}

We consider continuous-time quantum walks on the state space $X$,
starting from $0$ and driven by the unitary matrix
\begin{equation}\label{u_c}
  U_C(t):=e^{\sqrt{-1}t(I-P)}
  :=\sum_{i=0}^\infty\frac{(\sqrt{-1}t)^i}{i!}(I-P)^i,
  \quad t \ge 0,
\end{equation}
where $P$ is the transition probability matrix of the associated
random walk.

The {\it time-averaged distribution} of a continuous-time quantum
walk is defined by
\begin{align*}
  \bar{p}_C(x)
  &:=\lim_{T\to\infty}\frac{1}{T}\int_0^T\mathbf{P}(X_t=x|X_0=0)dt
  \nonumber\\
  &=\lim_{T\to\infty}\frac{1}{T}\int_0^T|\langle x|U_C(t)|0\rangle|^2dt
  =\lim_{T\to\infty}\frac{1}{T}\int_0^T|\{U_C(t)\}_{x,0}|^2dt,
  \quad x\in X.
\end{align*}
Since $\bar{p}_D(x)$ depends on $x$ through the Hamming distance
$|x|=\partial(0,x)$, the distribution of the Hamming distance is
given by
\[
  \bar{q}_C(h)=\sum_{y\in X:|y|=h}\bar{p}_C(y)
  =\begin{pmatrix}d\\h\end{pmatrix}\bar{p}_C(x),
  \quad h\in\{0,1,\ldots,d\},
\]
where $x$ satisfies $h=|x|$.

\begin{theorem}\label{thm:CTQW}
  Suppose the transition probability of a random walk satisfies
  the symmetry \eqref{assump} and that the eigenvalues \eqref{eigen}
  are distinct. Then the time-averaged distribution of the Hamming
  distance for the continuous-time quantum walk follows a discrete
  arcsine law.
\end{theorem}  

\begin{proof}
  By using the base change of eigenmatrices \eqref{eigenmat},
  we have
  \[
  (I-P)^i
  =\left\{I-\sum_{j=0}^d\frac{w_j}{\kappa_j}
  A_j\right\}^i
  =\left\{I-\sum_{j=0}^d\frac{w_j}{\kappa_j}
  \sum_{l=0}^d K_j(l)E_l\right\}^i
  =\sum_{l=0}^d(1-\rho_l)^iE_l,
  \]
  where the third equality follows since $E_0$, $E_1,\ldots$,
  $E_d$ are idempotent. Substituting this into \eqref{u_c},
  we have
  \[
  U_C(t)
  =\sum_{i=0}^\infty\frac{(\sqrt{-1}t)^i}{i!}
  \sum_{l=0}^d(1-\rho_l)^iE_l
  =\sum_{l=0}^d\sum_{i=0}^\infty\frac{(\sqrt{-1}t)^i}{i!}
  (1-\rho_l)^iE_l
  =\sum_{l=0}^de^{\sqrt{-1}(1-\rho_l)t}E_l.
  \]
  Then, by the base change, we have
  \begin{align*}
  \{U_C(t)\}_{x,0}&=\sum_{i=0}^d e^{\sqrt{-1}(1-\rho_i)t}(E_i)_{x0}
   =\sum_{i=0}^d e^{-\sqrt{-1}(\rho_i-1)t}
   \frac{1}{2^d}\sum_{l=0}^dK_i(l)(A_l)_{x0}\\
  &=\frac{1}{2^d}\sum_{i=0}^d e^{-\sqrt{-1}(\rho_i-1)t}
   \sum_{l=0}^dK_i(l)\delta_{l,|x|}
   =\frac{1}{2^d}\sum_{i=0}^d e^{-\sqrt{-1}(\rho_i-1)t}K_i(|x|).
  \end{align*}
  The squared modulus is
  \begin{align}
    \frac{1}{4^d}\sum_{i=0}^d\sum_{l=0}^d&
    e^{-\sqrt{-1}(\rho_i-\rho_l)t} K_i(|x|)K_l(|x|)\nonumber\\
    =&\frac{1}{4^d}\sum_{i=0}^d\{K_i(|x|)\}^2
    +\frac{2}{4^d}\sum_{i=0}^d\sum_{l\in\{0,1,\ldots,d\}\setminus\{i\}}
    \cos\{(\rho_i-\rho_l)t\} K_i(|x|)K_l(|x|)\nonumber\\
    =&\frac{\begin{pmatrix}2(d-|x|)\\d-|x|\end{pmatrix}
        \begin{pmatrix}2|x|\\|x|\end{pmatrix}}
         {4^d\begin{pmatrix}d\\|x|\end{pmatrix}}
     +\frac{2}{4^d}\sum_{i=0}^d\sum_{l\in\{0,1,\ldots,d\}\setminus\{i\}}
     \cos\{(\rho_i-\rho_l)t\} K_i(|x|)K_l(|x|),
    \label{modulus} 
  \end{align}
  where the last equality follows by \eqref{sum}.
  The first term is the probability mass function of a discrete
  arcsine law. The contribution of the second term vanishes, because
  \[
  \frac{1}{T}\int_0^T \cos\{(\rho_i-\rho_l)t\}K_i(|x|)K_l(|x|)dt=
  \frac{\sin\{(\rho_i-\rho_l)T\}}{T(\rho_i-\rho_l)}
  K_i(|x|)K_l(|x|)\to 0, \quad T\to\infty.  
  \]
  Therefore, we have
  \[
  \bar{p}_C(h)=\frac{1}{4^d}\begin{pmatrix}2(d-h)\\d-h\end{pmatrix}
        \begin{pmatrix}2h\\h\end{pmatrix}, \quad h\in\{0,1,\ldots d\},
  \]
  which is the probability mass function of the discrete arcsine law.
\end{proof}

This result covers the cases of quantum walks associated with
the simple random walk (Example~\ref{exa:simple}) and
the binomially weighted random walk (Example~\ref{exa:bin}).
The result for the simple random walk previously appeared in
\cite[Section~9.5]{Kon08}. On the other hand,
if some eigenvalues are degenerate, the time-averaged distribution
is not a discrete arcsine law. For the quantum walk associated
with the independent random walk (Example~\ref{exa:indep}),
the second term of the right-hand side of \eqref{modulus} becomes
\begin{align*}
  &\frac{2}{4^d}\cos t\sum_{l=1}^d(|x|)+
  \frac{1}{4^d}\left[\left\{\sum_{l=0}^dK_l(|x|)\right\}^2-\sum_{l=0}^d\{K_l(|x|)\}^2
  -2\sum_{l=1}^dK_l(|x|)\right]\\
  &=\frac{2}{4^d}\cos t(\delta_{|x|,0}2^d-1)
  -\frac{\begin{pmatrix}2(d-|x|)\\d-|x|\end{pmatrix}
      \begin{pmatrix}2|x|\\|x|\end{pmatrix}}
       {4^d\begin{pmatrix}d\\|x|\end{pmatrix}}
  +\delta_{|x|,0}\left(1-\frac{2}{2^d}\right)+\frac{2}{4^d},
\end{align*}  
where we used the generating function \eqref{gen} with $s=1$ and
\eqref{sum}. The time-averaged distribution is
\begin{equation}\label{ind_clim}
  \bar{p}_C(h)=\left(1-\frac{2}{2^d}\right)\delta_{h,0}+\frac{2}{4^d},
\end{equation}
which reproduces the result previously appeared in
\cite[Section~9.5]{Kon08}.

\section{Scaling limits of time-averaged distributions}\label{sec:limits}

In this section, we denote the probability mass function of
a discrete arcsine law as
\begin{equation}\label{dasl}
   q_d(h):=\frac{1}{4^d}\begin{pmatrix}2(d-h)\\d-h\end{pmatrix}
  \begin{pmatrix}2h\\h\end{pmatrix}, \quad h\in\{0,1,\ldots,d\}.
\end{equation}

\subsection{Continuous-time quantum walks}

According to Theorem~\ref{thm:CTQW}, if the transition probability
of a random walk satisfies the symmetry \eqref{assump} and the
eigenvalues in \eqref{eigen} are distinct, then the time-averaged
distribution of the Hamming distance $H$ for the continuous-time
quantum walk is the discrete arcsine law. It is well known that
if $H$ follows the discrete arcsine law, the rescaled variable
$\eta=H/d$ converges weakly to the arcsine law with density
\begin{equation}\label{arcsine}
  f(\eta):=\frac{1}{\pi}(1-\eta)^{-1/2}\eta^{-1/2}, \quad \eta\in(0,1).
\end{equation}
as $d\to\infty$. Examples include the continuous-time quantum walk
associated with the simple random walk (Example~\ref{exa:simple})
and the binomially weighted random walk (Example~\ref{exa:bin}).

\subsection{Discrete-time quantum walks}

Let $X=\{0,1\}^d$ be position and coin space,
respectively, equipped with Hilbert spaces with bases
$\{|x\rangle\}_{x\in X}$
and $\{|y\rangle\}_{y\in X}$. The dual bases are denoted by
$\{\langle x|\}_{x\in X}$ and $\{\langle y|\}_{y\in X}$.
A quantum state at step $t=0,1,\ldots$ is represented as
\[
  |\Psi(t)\rangle=
  \sum_{y\in X}\sum_{x\in X}\psi_{y,x}(t)|y,x\rangle,
  \quad |y,x\rangle=|y\rangle\otimes|x\rangle,
\]
where $\{\psi_{y,x}(t)\}\in\mathbb{C}^{X\times X}$ is
called the wave vector. The probability that we observe
the quantum walker at position $x\in X$ after $t$-steps is
\[
P_t(x)
=\sum_{y\in X}|\langle y,x|\Psi(t)\rangle|^2=\sum_{y\in X}|\psi_{y,x}(t)|^2.
\]
The discrete-time (coined) quantum walk is driven by
the evolution operator $U_D$:
\[
  |\Psi(t)\rangle = U^t_D|\Psi(0)\rangle
\]
where $U_D=S\circ(C\otimes I)$
and $C=\sum_{y,y'\in X}C_{y,y'}|y\rangle\langle y'|$ is called
a coin operator in $\mathcal{H}_C$, $I$ is the identity in
$\mathcal{H}_P$, and $S$ is the shift operator defined by
\[
  S=\sum_{y\in X}\sum_{x\in X}|y,x\oplus y\rangle\langle y,x|.
\]
Here, $x\oplus y$ is the component-wise sum of $d$-dimensional
vectors $x$ and $y$ of modulo $2$. As for the coin operator, we
take a common choice, so called Szegedy's walk \cite{Szegedy04},
associated with the random walk determined by the transition
probability of a random walk \eqref{trans}. In particular, the quantum walk
associated with the simple random walk is called the Grover walk.
Namely, \eqref{trans_1} determines
\begin{equation*}
  C_{y,y'}=2\sqrt{P_{0,y}P_{0,y'}}-\delta_{y,y'}
  =2\sqrt{\frac{w_{|y|}}{\kappa_{|y|}}\frac{w_{|y'|}}
  {\kappa_{|y'|}}}-\delta_{y,y'}, \quad y,y'\in X.
\end{equation*}
We set the initial state:
\begin{equation*}
  |\Psi(0)\rangle=\sum_{y\in X}\sum_{x\in X}\psi_{y,x}(0)|y,x\rangle
  =\sum_{y\in X}\sqrt{\frac{w_{|y|}}{\kappa_{|y|}}}|y,0\rangle,
\end{equation*}
which means that the quantum walk starts from $0$ with
the law \eqref{trans_1} in the coin space:
\[
  |\psi_{y,x}(0)|^2=\delta_{x,0}\frac{w_{|y|}}{\kappa_{|y|}},
  \quad x,y\in X.
\]
See \cite{GM26} for details.

The time-averaged distribution of discrete-time quantum walks is
defined as
\[
  \bar{p}^D_d(x)=\lim_{T\to \infty}\frac{1}{T}\sum_{t=0}^{T-1}P_t(x),
\]
and the distribution of the Hamming distance is given by
\[
  \bar{q}^D_d(h)=\begin{pmatrix}d\\h\end{pmatrix}\bar{p}^D_d(x),
  \quad h\in\{0,1,\ldots,d\},
\]
where $x$ satisfies $h=|x|$. The probability mass function of
the time-averaged distribution of the discrete-time quantum walk
associated with the simple random walk (Example~\ref{exa:simple})
has been obtained by several authors, including
\cite{MPAD08,HIKST18,GM26}. The corresponding distribution for
the binomially weighted random walk (Example~\ref{exa:bin})
was discussed in \cite{GM26}. In this subsection we derive
the scaling limits as $h,d\to\infty$ with $h/d\in(0,1)$.

\subsubsection{The quantum walk associated with the simple random walk}

The time-averaged distribution of the discrete-time quantum
walk associated with the simple random walk is
\begin{align}
  \bar{q}^D_d(h)
  =&\frac{q_d(h)}{2}+\frac{1}{4^d}\begin{pmatrix}d\\h\end{pmatrix}
  \nonumber\\
  &+\frac{2}{4^d}\begin{pmatrix}d\\h\end{pmatrix}
  (d-h)h\sum_{j=1}^{d-1}
  \frac{\{K_{j-1}(h-1;2,d-2)\}^2}{j(d-j)}1_{[1,d-1]}(h),
  \quad h\in\{0,1,\ldots,d\}, \label{sim_lim}
\end{align}
where $q_d(h)$ is the probability mass function of the discrete
arcsine law \eqref{dasl}. See \cite[Equation~13]{HIKST18} and
\cite[Example~6.1]{GM26} for derivations.

\begin{theorem}\label{thm:simp}
  If $H_d$ is a random variable following the time-averaged
  distribution of the Hamming distance for the discrete-time
  quantum walk associated with the simple random walk, then
  the rescaled variable $H_d/d$ converges weakly to
  the arcsine law with density~\eqref{arcsine} as $d\to\infty$.
\end{theorem}

\begin{proof}
Noting that
\[
  j(d-j)=\frac{d^2}{4}\left\{1-(1-2j/d)^2\right\},
\]
where $|1-2(j+1)/d|<1$ for $j\in\{0,1,\ldots,d-2\}$,
the sum in \eqref{sim_lim} is recast into
\begin{align}
  &\frac{1}{2\cdot 4^{d-2}}
  \begin{pmatrix}d\\h\end{pmatrix}
  \left(1-\frac{h}{d}\right)\frac{h}{d}
  \sum_{j=0}^{d-2}\{K_j(h-1;2,d-2)\}^2
  \left[1+\frac{\{1-2(j+1)/d\}^2}{1-\{1-2(j+1)/d\}^2}\right]1_{[1,d-1]}(h)
  \nonumber\\
  &=\frac{1}{2}\left(1-\frac{1}{d}\right)q_{d-2}(h-1)1_{[1,d-1]}(h)
  \nonumber\\
  &+\frac{1}{2}\left(1-\frac{1}{d}\right)\frac{1}{4^{d-2}}
  \begin{pmatrix}d-2\\h-1\end{pmatrix}
  \sum_{j=0}^{d-2}\{K_j(h-1;2,d-2)\}^2
  \frac{\{1-2(j+1)/d\}^2}{1-\{1-2(j+1)/d\}^2}1_{[1,d-1]}(h),
  \label{sim_sum}
\end{align}
where we used \eqref{sum}. The first term is the probability
mass function of the discrete arcsine law modulo $(1-1/d)/2$.

We will show that the remainder
\begin{align*}
  R_d(h)&=\frac{1}{4^d}\begin{pmatrix}d\\h\end{pmatrix}\\
  &+\frac{1}{2}\left(1-\frac{1}{d}\right)\frac{1}{4^{d-2}}
  \begin{pmatrix}d-2\\h-1\end{pmatrix}
  \sum_{j=0}^{d-2}\{K_j(h-1;2,d-2)\}^2
  \frac{\{1-2(j+1)/d\}^2}{1-\{1-2(j+1)/d\}^2}1_{[1,d-1]}(h)
\end{align*}
satisfies
\[
  \lim_{d\to\infty}\sum_{h=0}^d R_d(h)
  =2\lim_{d\to\infty}
  \left\|\bar{q}^D_d-\frac{1}{2}\left\{q_d+\left(1-\frac{1}{d}\right)q_{d-2}1_{[1,d-1]}\right\}
  \right\|_{\rm TV}=0,
\]
which implies the weak convergence of the law of the rescaled
variable $H_d/d$, where $H_d$ follows the probability
distribution with the probability mass function $\bar{q}^D_d(h)$,
to the arcsine law with the density \eqref{arcsine}, because
\begin{align*}
  \left|
  \sum_{h=0}^d g(h/d)\right.&\left.\bar{q}^D_d(h)
  -\int_0^1 g(\eta) f(\eta)d\eta\right|\\
  \le&\left|
  \sum_{h=0}^d g(h/d)\bar{q}^D_d(h)
  -\sum_{h=0}^d g(h/d)\frac{1}{2}\left\{q_d(h)
  +\left(1-\frac{1}{d}\right)q_{d-2}(h-1)1_{[1,d-1]}(h)\right\}\right|\\
  &+\left|\sum_{h=0}^d g(h/d)\frac{1}{2}\left\{q_d(h)
  +\left(1-\frac{1}{d}\right)q_{d-2}(h-1)1_{[1,h-1]}\right\}-\int_0^1 g(\eta) f(\eta)d\eta
  \right|\\
  \le&\sum_{h=0}^d |g(h/d)|\left|\bar{q}^D_d(h)
  -\frac{1}{2}\left\{q_d(h)+\left(1-\frac{1}{d}\right)q_{d-2}(h-1)1_{[1,d-1]}(h)\right\}\right|
  \\
  &+\left|\sum_{h=0}^d g(h/d)\frac{1}{2}\left\{q_d(h)
  +\left(1-\frac{1}{d}\right)q_{d-2}(h-1)1_{[1,h-1]}\right\}-\int_0^1 g(\eta) f(\eta)d\eta
  \right|\\
  \le& 2|g|_{\infty}\left\|\bar{q}^D_d
  -\frac{1}{2}\left\{q_d+\left(1-\frac{1}{d}\right)q_{d-2}1_{[1,d-1]}\right\}\right\|_{\rm TV}\\
  &+\left|\sum_{h=0}^d g(h/d)\frac{1}{2}\left\{q_d(h)
  +\left(1-\frac{1}{d}\right)q_{d-2}(h-1)1_{[1,h-1]}\right\}-\int_0^1 g(\eta) f(\eta)d\eta
  \right|
\end{align*}
for any bounded and continuous function $g$, where the last term in the
right-hand side tends to 0 as $d\to\infty$ by the weak convergence.
We have
\begin{align*}
  \sum_{h=0}^d R_d(h)&=\frac{1}{2^d}
  +\frac{1}{2}\left(1-\frac{1}{d}\right)\frac{1}{4^{d-2}}
  \sum_{j=0}^{d-2}\sum_{h=0}^{d-2}\begin{pmatrix}d-2\\h\end{pmatrix}
  \{K_j(h;2,d-2)\}^2
  \frac{\{1-2(j+1)/d\}^2}{1-\{1-2(j+1)/d\}^2}\\
  &=\frac{1}{2^d}
  +\frac{1}{2}\left(1-\frac{1}{d}\right)
  \sum_{j=0}^{d-2}
  \begin{pmatrix}d-2\\j\end{pmatrix}\frac{1}{2^{d-2}}
    \frac{\{1-2(j+1)/d\}^2}{1-\{1-2(j+1)/d\}^2}\\
  &=\frac{1}{2^d}
    +\frac{1}{2}\left(1-\frac{1}{d}\right)
  \mathbf{E}\left[  
  \frac{\{1-2(J+1)/d\}^2}{1-\{1-2(J+1)/d\}^2}\right]
\end{align*}
where we used the orthogonality \eqref{orth}, and the expectation
is with respect to the binomial distribution ${\rm Binom.}(d-2,1/2)$.
If $|j-(d-2)/2|<(d-2)^{(1+\epsilon)/2}$ for $\epsilon\in(0,1)$,
we have
\[
\frac{\{1-2(j+1)/d\}^2}{1-\{1-2(j+1)/d\}^2}
=\frac{\{j-(d-2)/2\}^2}{(d-j-1)(j+1)}=O(d^{-1+\epsilon})
\]
for large $d$. Therefore,
\begin{align*}
  0<\mathbf{E}\left[
    \frac{\{1-2(J+1)/d\}^2}{1-\{1-2(J+1)/d\}^2}\right]
  \le&
  O(d^{-1+\epsilon})
  \mathbf{P}\left\{|J-(d-2)/2|<(d-2)^{(1+\epsilon)/2}\right\}\\
  &+\frac{\{(d-2)/2\}^2}{d-1}
  \mathbf{P}\left\{|J-(d-2)/2|\ge(d-2)^{(1+\epsilon)/2}\right\}.
\end{align*}
Since Hoeffding's inequality gives
\[
\mathbf{P}\left\{|J-(d-2)/2|\ge(d-2)^{(1+\epsilon)/2}\right\}
\le2e^{-2(d-2)^\epsilon},
\]
we have $\sum_{h=0}^d R_d(h)=O(d^{-1+\epsilon})$ and we conclude
that $\lim_{d\to\infty}\sum_{h=0}^d R_d(h)=0$.
\end{proof}

\subsubsection{The quantum walk associated with the independent random walk}

The time-averaged distribution of the discrete-time quantum walk associated
with the independent random walk (Example~\ref{exa:indep}) is given by
\[
\bar{q}^D_d(|x|)=\left(\frac{1}{2}-\frac{1}{2^d}\right)\delta_{h,0}
+\frac{1}{2\cdot 2^d}\begin{pmatrix}d\\h\end{pmatrix}
+\frac{1}{4^d}\begin{pmatrix}d\\h\end{pmatrix}.
\]  
See \cite[Example~6.3]{GM26}. As $d\to\infty$,its scaling limit is
$(\delta_{\eta,0}+\delta_{\eta,1/2})/2$, which differs from
the continuous counterpart \eqref{ind_clim}.

\subsubsection{The quantum walk associated with the binomially weighted random walk}

The time-averaged distribution of the discrete-time quantum walk associated
with the binomially weighted random walk (Example~\ref{exa:bin}) is given
by
\begin{align*}
  \bar{q}^D_d(|x|)
    =&\frac{q_d(|x|)}{2}+\frac{1}{2\cdot 4^d}\begin{pmatrix}d\\|x|\end{pmatrix}
    \nonumber\\
    &+\frac{1}{2\cdot 4^d}\begin{pmatrix}d\\|x|\end{pmatrix}
    \sum_{y\in X}r^{|y|}(1-r)^{d-|y|}
    \sum_{j=1}^d
    \frac{\{(1-2r)^jK_j(|x|)-K_j(|x\oplus y|)\}^2}{1-(1-2r)^{2j}}\\
    =&\frac{q_d(|x|)}{2}+\frac{1}{2\cdot 4^d}\begin{pmatrix}d\\|x|\end{pmatrix}
    +\frac{1}{2\cdot 4^d}\begin{pmatrix}d\\|x|\end{pmatrix}\sum_{l=0}^d r^l(1-r)^{d-l}\\
    &\times\sum_{k=\max\{0,|x|+l-d\}}^{\min\{h,l\}}
    \begin{pmatrix}h\\k\end{pmatrix}
    \begin{pmatrix}d-|x|\\l-k\end{pmatrix}
    \sum_{j=1}^d
    \frac{\{(1-2r)^jK_j(|x|)-K_j(|x|-2k+l)\}^2}{1-(1-2r)^{2j}},
    \quad x\in\{0,1\}^d
\end{align*}
for $r\in(0,1)\setminus\{1/2\}$. This expression was obtained
in \cite[Equation~6.5]{GM26}, although there is a typographical
error in the denominator of the summand on the second line. We have
\begin{align}
&\bar{q}^D_d(h)
    =\frac{q_d(h)}{2}+\frac{1}{2\cdot 4^d}\begin{pmatrix}d\\h\end{pmatrix}
    \nonumber\\
    &+\frac{1}{2\cdot 4^d}\begin{pmatrix}d\\h\end{pmatrix}\sum_{l=0}^d r^l(1-r)^{d-l}\sum_{k=\max\{0,h+l-d\}}^{\min\{h,l\}}
    \begin{pmatrix}h\\k\end{pmatrix}
    \begin{pmatrix}d-h\\l-k\end{pmatrix}
    \sum_{j=1}^d
    \frac{\{(1-2r)^jK_j(h)-K_j(h-2k+l)\}^2}{1-(1-2r)^{2j}}\nonumber\\
    &=\frac{q_d(h)}{2}+\frac{1}{2\cdot 4^d}\begin{pmatrix}d\\h\end{pmatrix}
    \nonumber\\
    &+\frac{1}{2\cdot 4^d}\begin{pmatrix}d\\h\end{pmatrix}
    \sum_{k=0}^h
    \sum_{l=k}^{d-h+k} 
    \begin{pmatrix}h\\k\end{pmatrix}
    \begin{pmatrix}d-h\\l-k\end{pmatrix}r^l(1-r)^{d-l}
    \sum_{j=1}^d
    \frac{\{(1-2r)^jK_j(h)-K_j(h-2k+l)\}^2}{1-(1-2r)^{2j}}\nonumber\\
    &=\frac{q_d(h)}{2}+\frac{1}{2\cdot 4^d}\begin{pmatrix}d\\h\end{pmatrix}
    \nonumber\\
    &+\frac{1}{2\cdot 4^d}\begin{pmatrix}d\\h\end{pmatrix}
    \sum_{k=0}^h
    \sum_{l=0}^{d-h} 
    \begin{pmatrix}h\\k\end{pmatrix}
    \begin{pmatrix}d-h\\l\end{pmatrix}r^{l+k}(1-r)^{d-l-k}
    \sum_{j=1}^d
    \frac{\{(1-2r)^jK_j(h)-K_j(h-k+l)\}^2}{1-(1-2r)^{2j}}\nonumber\\
    &=\frac{q_d(h)}{2}+\frac{1}{2\cdot 4^d}\begin{pmatrix}d\\h\end{pmatrix}
    +\frac{1}{2\cdot 4^d}\begin{pmatrix}d\\h\end{pmatrix}
    \mathbf{E}\left[
      \sum_{j=1}^d
      \frac{\{(1-2r)^jK_j(h)-K_j(h-K+L)\}^2}{1-(1-2r)^{2j}}\right],
   \label{bin_lim} 
\end{align}  
where $K$ and $L$ independently follows the binomial distribution
${\rm Binom.}(h,r)$ and ${\rm Binom.}(d-h,r)$, respectively.

We prepare lemmas.

\begin{lemma}\label{lem:exp}
  If $K$ and $L$ independently follow the binomial distributions
  ${\rm Binom.}(h,r)$ and ${\rm Binom.}(d-h,r)$, respectively,
  then\[
    \mathbf{E}[K_j(h-K+L)]=(1-2r)^jK_j(h).
  \]
\end{lemma}

\begin{proof}
  Computing the generating function \eqref{gen} yields
  \begin{align*}
    \sum_{j=0}^d\mathbf{E}[&K_j(h-K+L)]s^j=
    \mathbf{E}\left[\sum_{j=0}^dK_j(h-K+L)s^j\right]\\
    &=(1-r)^d(1+s)^{d-h}(1-s)^h
    \sum_{k=0}^h\sum_{l=0}^{d-h}
    \begin{pmatrix}h\\k\end{pmatrix}
    \begin{pmatrix}d-h\\l\end{pmatrix}
    \left(\frac{r}{1-r}\frac{1+s}{1-s}\right)^k
    \left(\frac{r}{1-r}\frac{1-s}{1+s}\right)^l\\
  &=(1-r)^d(1+s)^{d-h}(1-s)^{h}
    \left(1+\frac{r}{1-r}\frac{1+s}{1-s}\right)^h
    \left(1+\frac{r}{1-r}\frac{1-s}{1+s}\right)^{d-h}\\
  &=\{1+(1-2r)s\}^{d-h}\{1-(1-2r)s\}^h
  =\sum_{j=0}^d K_j(h)\{(1-2r)s\}^j.
  \end{align*}
  Comparing the coefficients of powers of $s$ establishes
  the assertion.
\end{proof}

\begin{lemma}\label{lem:conv}
  Let
  \begin{align*}
  M_d(h)&=\frac{1}{4^d}\begin{pmatrix}d\\h\end{pmatrix}
    \mathbf{E}\left[
      \frac{
        \begin{pmatrix}2(d-h+K-L)\\d-h+K-L\end{pmatrix}  
        \begin{pmatrix}2(h-K+L)\\h-K+L\end{pmatrix}}  
           {\begin{pmatrix}d\\h-K+L\end{pmatrix}}\right],
  \quad h\in\{0,1,\ldots,d\}  
  \end{align*}
  where $K$ and $L$ independently follow the binomial distributions
  ${\rm Binom.}(h,r)$ and ${\rm Binom.}(d-h,r)$, respectively.
  Then, $M_d(h)$ is a probability mass function. Moreover, if $H_d$
  is a random variable with the probability mass function $M_d(h)$,
  then the rescaled variable $H_d/d$ converges weakly as
  $d \to \infty$ to the law of $r+(1-2r)\xi$, where $\xi$ follows the
  arcsine law on $(0,1)$.
\end{lemma}

\begin{proof}
  We have
  \begin{align}
    M_d(h)&=\sum_{k=0}^h\sum_{l=0}^{d-h}
    \begin{pmatrix}h-k+l\\l\end{pmatrix}
    \begin{pmatrix}d-h+k-l\\k\end{pmatrix}
      r^{k+l}(1-r)^{d-k-l}
      q_d(h-k+l)\nonumber\\
  &=\sum_{i=0}^h\sum_{j=i}^{d-h+i}
    \begin{pmatrix}j\\i\end{pmatrix}
    \begin{pmatrix}d-j\\h-i\end{pmatrix}
      r^{h+j-2i}(1-r)^{d-h-j+2i}q_d(j)\nonumber\\
  &=\sum_{j=0}^d\sum_{i=\max\{0,j-d+h\}}^{\min\{h,j\}}
    \begin{pmatrix}j\\i\end{pmatrix}
    \begin{pmatrix}d-j\\h-i\end{pmatrix}
      r^{h+j-2i}(1-r)^{d-h-j+2i}q_d(j),\label{mix}      
  \end{align}  
  where we set $i=h-k$, $j=h-k+l$, and $q_d(j)$ is the probability
  mass function of the discrete arcsine law \eqref{dasl}.
  Since $q_d(j)$ is a probability mass function,
  if the summand is a probability mass function for fixed $j$,
  then $M_d(h)$ is a probability mass function of a mixture
  distribution. In fact, since
  \begin{align*}
    &\sum_{h=0}^d\sum_{i=\max\{0,j-d+h\}}^{\min\{h,j\}}
    \begin{pmatrix}j\\i\end{pmatrix}
    \begin{pmatrix}d-j\\h-i\end{pmatrix}
      r^{h+j-2i}(1-r)^{d-h-j+2i}\\
    &=\sum_{i=0}^j\sum_{l=0}^{d-j}
    \begin{pmatrix}j\\i\end{pmatrix}
    \begin{pmatrix}d-j\\l\end{pmatrix}
      r^{j-i+l}(1-r)^{d-j+i-l}\\
      &=r^j(1-r)^{d-j}\sum_{i=0}^j
      \begin{pmatrix}j\\i\end{pmatrix}
        \left(\frac{1-r}{r}\right)^i\sum_{l=0}^{d-j}
    \begin{pmatrix}d-j\\l\end{pmatrix}
      \left(\frac{r}{1-r}\right)^l=1,
  \end{align*}    
  $M_d(h)$ is a probability mass function, where we set $l=h-i$.
  We evaluate the characteristic function of $H_d/d$, where $H_d$
  is a random variable with the probability mass function $M_d(h)$.
  \begin{align*}
  \varphi_d(t)&:=\sum_{h=0}^de^{\sqrt{-1}th/d}M_d(h)\\
  &=\sum_{j=0}^d\sum_{i=0}^j
   \begin{pmatrix}j\\i\end{pmatrix}
   \sum_{l=0}^{d-j} 
   \begin{pmatrix}d-j\\l\end{pmatrix}
   r^{j-i+l}(1-r)^{d-j+i-l}q_d(j)e^{\sqrt{-1}t(l+i)/d}\\
   &=\sum_{j=0}^d\sum_{i=0}^j
   \begin{pmatrix}j\\i\end{pmatrix}
   \left(\frac{1-r}{r}e^{\sqrt{-1}t/d}\right)^i  
   \sum_{l=0}^{d-j} 
   \begin{pmatrix}d-j\\l\end{pmatrix}
   \left(\frac{r}{1-r}e^{\sqrt{-1}t/d}\right)^l
   r^{j}(1-r)^{d-j}q_d(j)\\
   &=\sum_{j=0}^d\left(1+\frac{1-r}{r}e^{\sqrt{-1}t/d}\right)^j  
   \left(1+\frac{r}{1-r}e^{\sqrt{-1}t/d}\right)^{d-j}
   r^{j}(1-r)^{d-j}q_d(j)\\
   &=\sum_{j=0}^d\left(r+(1-r)e^{\sqrt{-1}t/d}\right)^j  
   \left((1-r)+re^{\sqrt{-1}t/d}\right)^{d-j}q_d(j),
  \end{align*}
  which is continuous at $t=0$ for all $d$. For large $d$,
  \[
    \varphi_d(t)=\mathbf{E}
    \left[e^{\sqrt{-1}t\{r+(1-2r)J/d\}}\left\{1+O(d^{-1})\right\}\right],
  \]
  where the expectation is with respect to the discrete arcsine law.
  Since $H_d/d$ converges weakly to the weak limit of $r+(1-2r)J/d$,
  which is the law of $r+(1-2r)\eta$, where $\eta$ follows
  the arcsine law on $(0,1)$.
\end{proof}

\begin{theorem}\label{thm:bin}
  If $H_d$ is a random variable following the time-averaged
  distribution of the Hamming distance for the discrete-time quantum
  walk associated with the binomially weighted random walk
  (Example~\ref{exa:bin}) with $r\in(0,1)\setminus\{1/2\}$, then
  the rescaled variable $H_d/d$ converges weakly as $d\to\infty$
  to the equal mixture of the arcsine law on $(0,1)$ and the law
  of $r+(1-2r)\eta$, where $\eta$ independently follows the arcsine
  law.
\end{theorem}

\begin{proof}
  By \eqref{bin_lim} and Lemma~\ref{lem:exp}, we have
  \begin{align*}
  \bar{q}^D_d(h)
    =&\frac{q_d(h)}{2}+\frac{1}{2\cdot 4^d}\begin{pmatrix}d\\h\end{pmatrix}
    +\frac{1}{2\cdot 4^d}\begin{pmatrix}d\\h\end{pmatrix}
    \left[
      \sum_{j=1}^d
      \frac{\mathbf{E}[\{K_j(h-K+L)\}^2]-\{\mathbf{E}\{K_j(h-K+L)\}\}^2}
           {1-(1-2r)^{2j}}\right]\\
    =&\frac{q_d(h)}{2}+\frac{1}{2\cdot 4^d}\begin{pmatrix}d\\h\end{pmatrix}
    +\frac{1}{2\cdot 4^d}\begin{pmatrix}d\\h\end{pmatrix}
    \left[
      \sum_{j=1}^d
      \frac{\mathbf{E}[\{K_j(h-K+L)\}^2]-(1-2r)^{2j}\{K_j(h)\}^2}
           {1-(1-2r)^{2j}}\right]\\
    =&\frac{q_d(h)}{2}+\frac{1}{2\cdot 4^d}\begin{pmatrix}d\\h\end{pmatrix}
      +\frac{1}{2\cdot 4^d}\begin{pmatrix}d\\h\end{pmatrix}
         \sum_{j=1}^d\mathbf{E}[\{K_j(h-K+L)\}^2]\\
     &+\frac{1}{2\cdot 4^d}
      \sum_{j=1}^d\left\{\mathbf{E}[\{K_j(h-K+L)\}^2]-\{K_j(h)\}^2
      \right\}
      \frac{(1-2r)^{2j}}{1-(1-2r)^{2j}}\\
      =&\frac{q_d(h)+M_d(h)}{2}
      +\frac{1}{2\cdot 4^d}\begin{pmatrix}d\\h\end{pmatrix}\\
      &+\frac{1}{2\cdot 4^d}\begin{pmatrix}d\\h\end{pmatrix}
      \sum_{j=1}^d\left\{\mathbf{E}[\{K_j(h-K+L)\}^2]-\{K_j(h)\}^2
      \right\}
      \frac{(1-2r)^{2j}}{1-(1-2r)^{2j}},  
  \end{align*}
  where we used \eqref{sum} in the last equality, and $M_d(h)$ is
  defined in Lemma~\ref{lem:conv}.

  We will show that the reminder:
  \[
      S_d(h)=\frac{1}{2\cdot 4^d}\begin{pmatrix}d\\h\end{pmatrix}
      +\frac{1}{2\cdot 4^d}\begin{pmatrix}d\\h\end{pmatrix}
      \sum_{j=1}^d\left\{\mathbf{E}[\{K_j(h-K+L)\}^2]-\{K_j(h)\}^2
      \right\}\frac{(1-2r)^{2j}}{1-(1-2r)^{2j}},  
  \]
  satisfies $\lim_{d\to\infty}\sum_{h=0}^d|S_d(h)|=0$.
  As in the proof of Theorem~\ref{thm:simp},
  with Lemma~\ref{lem:conv}, this convergence implies the weak
  convergence of the law of the rescaled variable $H_d/d$, where
  $H_d$ is a random variable with the probability mass function
  $\bar{q}^D_d(h)$, to the equal mixture of the arcsine law on
  $(0,1)$ and the law of $r+(1-2r)\eta$, where $\eta$ independently
  follows the arcsine law. We have
  \begin{align}
    2\sum_{h=0}^d|S_d(h)|\le&
    \frac{1}{4^d}\sum_{h=0}^d\begin{pmatrix}d\\h\end{pmatrix}
    +\frac{1}{4^d}\sum_{j=1}^d\sum_{h=0}^d\begin{pmatrix}d\\h\end{pmatrix}
    \{K_j(h)\}^2\frac{(1-2r)^{2j}}{1-(1-2r)^{2j}}\nonumber\\
    &+\frac{1}{4^d}
    \sum_{j=1}^d\sum_{h=0}^d\begin{pmatrix}d\\h\end{pmatrix}
    \mathbf{E}[\{K_j(h-K+L)\}^2]\frac{(1-2r)^{2j}}{1-(1-2r)^{2j}}\nonumber\\
    =&\frac{1}{2^d}
    +\frac{1}{2^d}
    \sum_{j=1}^d\begin{pmatrix}d\\j\end{pmatrix}
    \frac{(1-2r)^{2j}}{1-(1-2r)^{2j}}\nonumber\\
    &+\frac{1}{4^d}
    \sum_{j=1}^d\sum_{h=0}^d\begin{pmatrix}d\\h\end{pmatrix}
    \mathbf{E}[\{K_j(h-K+L)\}^2]\frac{(1-2r)^{2j}}{1-(1-2r)^{2j}},
    \label{bin_sum}
  \end{align}
  where in the equality we used the orthogonality \eqref{orth}.
  The first term tends to 0 as $d\to\infty$. The second term also
  tends to 0, because
  \begin{align*}
    \frac{1}{2^d}
    \sum_{j=1}^d\begin{pmatrix}d\\j\end{pmatrix}
    \frac{(1-2r)^{2j}}{1-(1-2r)^{2j}}
    &\le 
    \frac{1}{2^d}\frac{1}{1-(1-2r)^{2}}
    \sum_{j=1}^d\begin{pmatrix}d\\j\end{pmatrix}(1-2r)^{2j}\\
    &= 
    \frac{1}{1-(1-2r)^{2}}\frac{\{1+(1-2r)^2\}^d-1}{2^d}\to 0,
    \quad d\to\infty.
  \end{align*}
  For the last term in \eqref{bin_sum}, we note that
  \begin{align*}
    \mathbf{E}[\{K_j(h-K+L)\}^2]
    &=\sum_{i=0}^h\sum_{l=0}^{d-h}
    \begin{pmatrix}h\\i\end{pmatrix}(1-r)^ir^{h-i}
    \begin{pmatrix}d-h\\l\end{pmatrix}r^l(1-r)^{d-h-l}\{K_j(i+l)\}^2\\
    &=\sum_{i=0}^h\sum_{m=i}^{d-h+i}
      \begin{pmatrix}h\\i
      \end{pmatrix}
      \begin{pmatrix}d-h\\m-i
      \end{pmatrix}r^{m+h-2i}(1-r)^{d-m-h+2i}\{K_j(m)\}^2,
  \end{align*}
  where we set $m=i+l$. Then, by changing the order of the summations,
  we have
  \begin{align*}
    &\sum_{h=0}^d\begin{pmatrix}d\\h\end{pmatrix}\mathbf{E}[\{K_j(h-K+L)\}^2]\\
    &=\sum_{h=0}^d\sum_{i=0}^h\sum_{m=i}^{d-h+i}
    \begin{pmatrix}d\\h\end{pmatrix}
    \begin{pmatrix}h\\i\end{pmatrix}
    \begin{pmatrix}d-h\\m-i\end{pmatrix}
    r^{m+h-2i}(1-r)^{d-m-h+2i}\{K_j(m)\}^2\\
    &=\sum_{i=0}^d\sum_{m=i}^d\sum_{h=i}^{d-m+i}
    \begin{pmatrix}d\\h\end{pmatrix}
    \begin{pmatrix}h\\i\end{pmatrix}
    \begin{pmatrix}d-h\\m-i\end{pmatrix}
    r^{m+h-2i}(1-r)^{d-m-h+2i}\{K_j(m)\}^2\\
    &=\sum_{i=0}^d\sum_{m=i}^d
    \begin{pmatrix}d\\m\end{pmatrix}
    \begin{pmatrix}m\\i\end{pmatrix}r^{m-i}(1-r)^i\{K_j(m)\}^2
    \sum_{k=0}^{d-m}\begin{pmatrix}d-m\\k\end{pmatrix}
      r^{k}(1-r)^{d-m-k}\\
    &=\sum_{i=0}^d\sum_{m=i}^d
    \begin{pmatrix}d\\m\end{pmatrix}
    \begin{pmatrix}m\\i\end{pmatrix}r^{m-i}(1-r)^i\{K_j(m)\}^2\\
    &=\sum_{m=0}^d
    \begin{pmatrix}d\\m\end{pmatrix}\{K_j(m)\}^2
    \sum_{i=0}^m\begin{pmatrix}m\\i\end{pmatrix}r^{m-i}(1-r)^i
    =\sum_{m=0}^d\begin{pmatrix}d\\m\end{pmatrix}\{K_j(m)\}^2
    =2^d\begin{pmatrix}d\\j\end{pmatrix}.
  \end{align*}
  Therefore, the last term of \eqref{bin_sum} tends to 0
  as the second term, and the assertion follows.
\end{proof}

\section{Discussion}\label{sec:disc}

Ide \cite{Ide23} showed that for quantum walks on path
graphs with one-step transitions, if the spectral gap
persists as the size of the graph grows, then the scaling
limits of the time-averaged distributions of
the continuous-time and discrete-time models coincide.
Our examples lie outside this setting and show that
the spectral gap alone does not decide the matter, in
either direction.

For the simple random walk on the hypercube, the gap
closes as $d\to\infty$, yet the two limits coincide
(the arcsine law). So the gap is not necessary.
For the random walk with binomial weights and for
the independent random walk, the gap persists, yet
the two limits differ. So the gap is not sufficient.

It is therefore an open problem to find a necessary and
sufficient condition for the coincidence of the scaling
limits. We regard the present examples as a first step
in this direction.
    
\section*{Acknowledgements}

Shuhei Mano was supported in part by JSPS KAKENHI Grant 24K06876.
The authors thank Robert Griffiths for comments.



\begin{flushleft}

\smallskip

Yusuke Ide\\
Department of Mathematics,
College of Humanities and Sciences,
Nihon University, Setagaya-ku, Tokyo, 156-8550, Japan\\
E-mail: ide.yusuke@nihon-u.ac.jp

\smallskip

Norio Konno\\
Department of Mathematical Sciences,
College of Science and Engineering,
Ritsumeikan University, Kusatsu, Shiga 525-8577, Japan\\
E-mail: n-konno@fc.ritsumei.ac.jp

\smallskip

Shuhei Mano\\
The Institute of Statistical Mathematics, Tachikawa, Tokyo 190-8562, Japan\\
E-mail: smano@ism.ac.jp

\end{flushleft}

\end{document}